%% file: template.tex
\documentclass[cameraready]{Interspeech}
\usepackage{multirow}
\usepackage{booktabs}   
\usepackage{tabularx}   
\usepackage{makecell}   
\usepackage{array}   
\usepackage{amsmath}
\title{A Unified and Reproducible Experimentation Framework for Speech Understanding}

\author[affiliation={1}, equalcontribution]{Jing}{Peng}
\author[affiliation={2}, equalcontribution]{Junhao}{Du}
\author[affiliation={2}, equalcontribution]{Chenghao}{Wang}
\author[affiliation={1}, equalcontribution]{Hanqi}{Li}
\author[affiliation={1}, equalcontribution]{Yi}{Yang}
\author[affiliation={2}]{Yixuan}{Wang}
\author[affiliation={1}]{Xiaoyu}{Gu}
\author[affiliation={1}]{Guanyu}{Chen}
\author[affiliation={3}]{Yucheng}{Wang}
\author[affiliation={5}]{Jiang}{Li}
\author[affiliation={5}]{Zhangjie}{Zhao}
\author[affiliation={1}]{Haoran}{Wang}
\author[affiliation={1}]{Wenming}{Tu}
\author[affiliation={1}]{Haoyu}{Li}
\author[affiliation={6}]{Duo}{Ma}
\author[affiliation={1}]{Lirong}{Qian}
\author[affiliation={1}]{Yu}{Xi}
\author[affiliation={1}]{Wen}{Wen}
\author[affiliation={2}]{Jiaqi}{Guo}
\author[affiliation={2}]{Hui}{Zhang}
\author[affiliation={2}]{Shuai}{Fan}
\author[affiliation={5}]{Wenbin}{Jiang}
\author[affiliation={4}]{Shuai}{Wang}
\author[affiliation={1}, correspondingauthor]{Kai}{Yu}

\address{
  $^1$X-LANCE Lab, Department of Computer Science and Engineering, Shanghai Jiao Tong University\\
  $^1$MoE Key Lab of Artificial Intelligence,
  $^1$Jiangsu Key Lab of Language Computing, China\\
  $^2$AISpeech Ltd, Suzhou, China
  $^3$ETH Zürich, Switzerland
  $^4$Nanjing University, Suzhou, China\\
  $^5$Hangzhou Dianzi University, Hangzhou, China
  $^6$The Chinese University of Hong Kong, Shenzhen, China
}

\email{\{jing.peng, kai.yu\}@sjtu.edu.cn}


\usepackage{comment}

\begin{document}

\maketitle

\begin{abstract}
Speech foundation models and Speech LLMs have advanced speech understanding, yet deployment-oriented model selection is hindered by non-comparable evaluations caused by mismatched post-processing, and by training results that are hard to reproduce across data scales and pipelines. We present \textbf{SURE}, a unified experimentation framework that standardizes prediction formats, normalization, and scoring. SURE evaluates strong systems across paradigms, from conventional pipelines to Speech LLMs, on representative tasks under realistic acoustic and linguistic stressors. Beyond evaluation, SURE introduces an agent-assisted training conversion flow that maps paper and code into versioned, runnable training pipelines under a unified protocol on matched open-data subsets. Overall, SURE improves comparability and reproducibility for deployment-oriented evaluation.
\end{abstract}

\input{src/introduction}
\input{src/overview}
\begin{figure}[htbp]
    \centering
    \includegraphics[width=0.95\columnwidth, ]{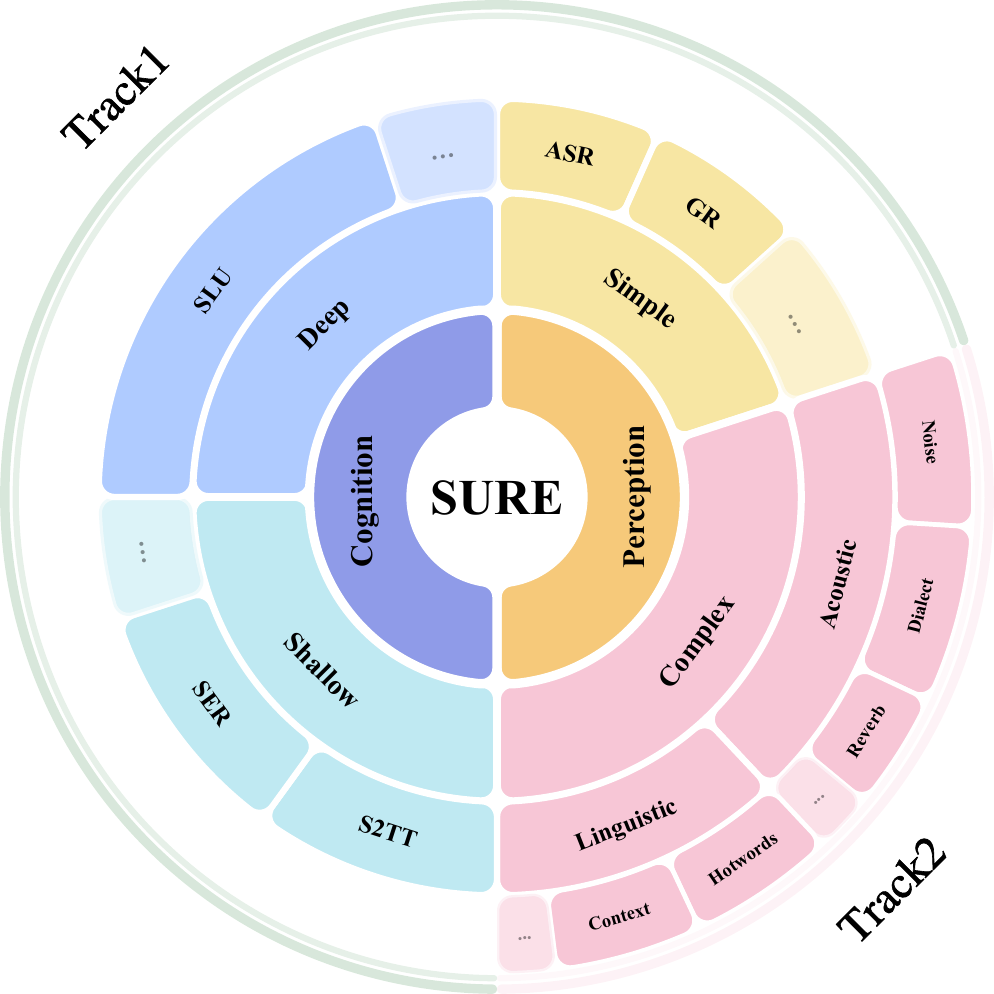}
    \caption{SURE Framework Evaluation Scope.}
    \label{fig:sure_scope}
\end{figure}

\begin{figure*}[htbp]  
    \centering
    \includegraphics[width=\textwidth, height=0.32\textheight]{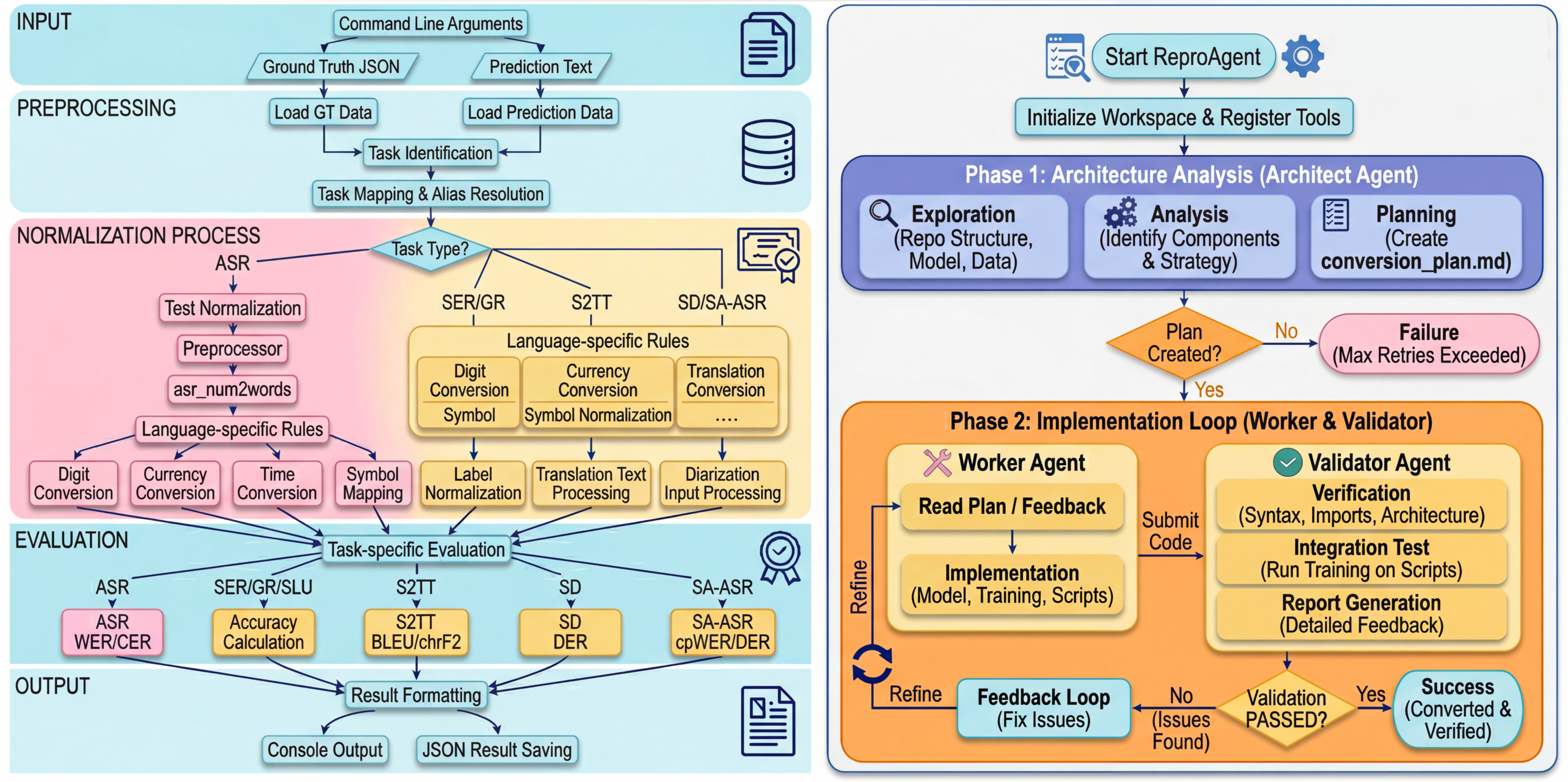}
    \caption{Overview of the SURE framework. Left: the unified evaluation pipeline with standardized prediction formats, post-processing, and scoring. Right:the agent-assisted training conversion workflow that maps ``paper + code'' into \textsc{swift} training recipes.}
    \label{fig:sure_structure}
\end{figure*}



\input{src/metrics}

\input{src/tracks}
\input{src/conclusion}
\newpage
\bibliographystyle{IEEEtran}
\bibliography{mybib}

\end{document}

%% file: src/introduction.tex
\section{Introduction}
\label{sec:intro}

Speech understanding has advanced rapidly with the rise of speech foundation models and speech large language models (Speech LLMs) \cite{arora2025landscapespokenlanguagemodels,Peng_2025}. These models aim to cover broad speech understanding capabilities, from recognition to understanding and reasoning over spoken content. Despite this progress, the community still lacks a reproducible framework that is suitable for model selection in both research and production. Reported results across papers and products are often \textbf{not directly comparable} due to \emph{inconsistent scoring pipelines, limited coverage of real-world conditions,large differences in training data scale and training pipelines}.

A central challenge is the \textbf{lack of standardization in evaluation}. Small choices in post-processing and scoring, such as text normalization, casing and punctuation handling, label mapping, segmentation, and task-specific heuristics, can materially change the final metrics~\cite{10.1145/3636513,srivastav2025openasrleaderboardreproducible,speakerdiarization}. This weakens the credibility of reported numbers and increases the cost of reproduction.

A second challenge is the \textbf{lack of generality in benchmarking}. Existing evaluations often cover only a narrow set of systems or capabilities~\cite{arora2024evaluationspeechfoundationmodels,yang2025speechrbenchmarkspeechreasoning}, which hinders cross-paradigm comparison and leaves key dimensions under-measured. A practical benchmark should \emph{cover strong-performing systems across paradigms}, not limited to Speech LLMs, but also including conventional models. It should also measure the \emph{broad capabilities} of speech understanding, and stress-test robustness under \emph{challenging conditions}.

A third challenge is the \textbf{lack of reproducible and fair architectural comparison}. Since modern Speech LLMs are trained with heterogeneous data mixtures and scales, gains are difficult to attribute to modeling choices rather than training conditions. This calls for ~\emph{matched open data, a unified training framework, and from-scratch recipes} that enable reproducible, architecture-level comparisons. We refer to this as \textbf{controlled training}, which standardizes data and training pipelines to reduce implementation variance.

To address these challenges, we introduce \textbf{SURE: A Unified and Reproducible Experimentation Framework for Speech Understanding}. Rather than a static benchmark, SURE targets \emph{deployment-oriented model selection} by coupling (i) scenario-driven evaluation under a fixed protocol, (ii) broad coverage spanning diverse model families and rich evaluation axes, and (iii) controlled training with an agent-assisted conversion workflow for fairer architectural comparison.

Recent benchmarks such as SUPERB, Dynamic SUPERB, AIR-Bench, and AudioBench~\cite{yang21c_interspeech,huang2024dynamic_superb,yang2024air_bench,wang2025audiobench} have greatly expanded task coverage and promoted large-scale evaluations. However, as summarized in Table~\ref{tab:benchmark_comparison}, these benchmarks typically provide limited model-family breadth within speech understanding tasks: \textit{evaluations are often centered on a narrow slice of model families}, typically Speech LLMs, with fewer side-by-side comparisons against strong conventional paradigms. For several tasks, Speech LLMs are not necessarily the best-performing system class, and the lack of architectural diversity prevents meaningful within-task conclusions. Moreover, even when a model is evaluated, \emph{results are usually reported under a single canonical test condition}, making it hard to precisely position the model’s robustness and generality under realistic stressors. In contrast, SURE emphasizes scenario-deep evaluation: probing each model across realistic stressors. It also supports multi-type models evaluation under unified protocols and controlled training.


The remainder of this paper is organized as follows. The Section~\ref{sec:overview} summarizes the released SURE package, including its tracks, data suites, and the unified interface for training, inference, and scoring. Section~\ref{sec:eval_metrics} then details the evaluation protocol and metrics. We report results in three tracks covering scenario-deep evaluation, cross-task comparison, and controlled training-based architecture studies, as detailed in Section~\ref{sec:track1}, Section~\ref{sec:track2}, and Section~\ref{sec:track3}, respectively. Our contributions are threefold:
\begin{itemize}
  \item We release \textsc{SURE} as a reproducible experimentation loop for deployment-oriented model selection, unifying evaluation under a consistent protocol within the framework.
  \item We curate scenario-focused suites and benchmark diverse model families and capability axes to enable cross-paradigm comparison under realistic acoustic and linguistic stressors.
  \item We take an initial step toward more controlled comparisons by introducing an agent-assisted conversion pipeline for controlled training that reduces implementation variance.
\end{itemize}

%% file: src/overview.tex
\section{Overview of SURE}
\label{sec:overview}


\newcolumntype{Y}{>{\centering\arraybackslash}X}
\begin{table}[t]
\centering
\caption{Comparison of speech understanding benchmarks in terms of \textbf{generality}. We report dataset stressors, model-family breadth, and controlled training for structure evaluation.}
\vspace{-2mm}
{\footnotesize \emph{Families} counts the number of distinct modeling paradigms that are explicitly evaluated side-by-side in each benchmark's main results (e.g., CTC/AED, cascaded pipelines, and Speech LLMs).}
\label{tab:benchmark_comparison}
\footnotesize
\setlength{\tabcolsep}{4pt}
\renewcommand{\arraystretch}{1.12}
\begin{tabularx}{\columnwidth}{l @{\hspace{2pt}} Y Y Y Y}
\toprule
\textbf{Benchmark} &
\multicolumn{2}{c}{\textbf{Datasets}} &
\textbf{Structure} &
\textbf{Models} \\
& \textbf{Acoustic} & \textbf{Linguistic} & \textbf{Evaluation} & \textbf{Families} \\
\midrule
SUPERB~\cite{yang2021superbspeechprocessinguniversal} & \checkmark & $\times$ & $\times$ & 1 \\
Dynamic~\cite{huang2024dynamic_superb}               & \checkmark & $\times$ & $\times$ & 2 \\
MMAU~\cite{sakshi2024mmaumassivemultitaskaudio}      & $\times$   & \checkmark & $\times$ & 1 \\
MMAR~\cite{ma2025mmarchallengingbenchmarkdeep}       & $\times$   & \checkmark & $\times$ & 1 \\
\midrule
\textbf{SURE (Ours)}                                 & \textbf{\checkmark} & \textbf{\checkmark} & \textbf{\checkmark} & \textbf{4} \\
\bottomrule
\end{tabularx}
\end{table}

SURE is an end-to-end experimentation package for reproducible speech understanding evaluation, with an agent-assisted conversion workflow for controlled training studies. It provides:

\begin{itemize}
  \item \textbf{Project website} for documentation and updates.\footnote{\ \textcolor{gray}{\url{https://sure-eval-framework.github.io/speechllm_series/}}}
  \item \textbf{Unified evaluation stack} with post-processing and scoring.\footnote{\ \textcolor{gray}{\url{https://anonymous.4open.science/r/evaluation-pipeline-839C}}}
  \item \textbf{Agent assisted pipeline} for training conversion .\footnote{\ \textcolor{gray}{\url{https://anonymous.4open.science/r/ReproAgent-9898}}}
  \item \textbf{Test and train suites} for all tracks.\footnote{\ \textcolor{gray}{\url{https://modelscope.cn/datasets/SUREBenchmark/SURE_Test_Suites}}}
\end{itemize}

SURE comprises three tracks: two for \emph{evaluation} (Track~I--II; Fig.~\ref{fig:sure_scope}) and one for \emph{controlled training studies} (Track~III).

\begin{itemize}
  \item \textbf{Track~I: Scenario Stress Testing for Front-end Speech Tasks.}
  Scenario suites curated from open-source data to \emph{approximate} real-world conditions, covering acoustic and linguistic stressors.

  \item \textbf{Track~II: Full-stack Speech Understanding Evaluation.}
  Unified evaluation of broad capabilities, from signal-level perception to lightweight semantic processing and transformation, and further to information-grounded deep reasoning.

  \item \textbf{Track~III: Initial Exploration of Controlled Training.}
  An initial step toward more controlled comparisons via from-scratch training on matched open-data subsets, enabled by an agent-assisted workflow that converts ``paper + code'' into runnable \texttt{swift} pipelines.
\end{itemize}

%% file: src/metrics.tex
\section{Evaluation Protocol and Metrics}
\label{sec:eval_metrics}

To connect the track design in Section~\ref{sec:overview} with the reproducibility goal stated in Section~\ref{sec:intro}, we summarize SURE's evaluation protocol and metrics. To enable reproducible evaluation across the three tracks, we release a unified evaluation pipeline with task-specific scoring rules, and further introduce RPS---a compact indicator for intuitive and updatable model comparison across heterogeneous speech-understanding tasks\cite{Peng_2025}.

\noindent\textbf{Unified evaluation pipeline.}
As shown in Fig.~\ref{fig:sure_structure} (left), our evaluation pipeline follows a fixed input--preprocess--normalize--score--report workflow.
Given a ground-truth JSON and a prediction text file, SURE first performs \emph{task identification} and \emph{alias resolution} to map user-specified task names to canonical evaluators. %
It then materializes per-task reference/hypothesis pairs into temporary files under a unified interface and invokes the corresponding scorer. %
For SD and SA-ASR, we use the official \texttt{meeteval}\cite{neumann2023meeteval} backends to compute DER and cpWER/DER on RTTM/STM-style inputs with a configurable collar. %
For S2TT, we compute BLEU and chrF2 using \texttt{sacrebleu}\cite{post-2018-call}. %
For text-centric tasks, we apply fixed normalization before scoring, including language-dependent number normalization and tag removal when applicable.
All task results are aggregated and saved in a unified JSON report.

\noindent\textbf{RPS as a dynamic and extensible indicator.}
Since SURE covers heterogeneous tasks and metrics, we report a unified \textbf{Relative Performance Score (RPS)} in $[0,1]$ by normalizing each task metric against the current best score on the \emph{SURE leaderboard} under the same evaluation pipeline, denoted as $\mathrm{Best}_t$:
\begin{equation}
\begin{aligned}
\mathrm{RPS}_t &=
\begin{cases}
\dfrac{s_t}{\mathrm{Best}_t+\epsilon}, & \text{higher-is-better},\\[2pt]
\dfrac{\mathrm{Best}_t}{s_t+\epsilon}, & \text{lower-is-better},
\end{cases}\\
\mathrm{RPS}_t &\leftarrow \min(\mathrm{RPS}_t, 1).
\end{aligned}
\end{equation}
where $s_t$ is the model score on task $t$ and $\epsilon$ is a small constant for numerical stability.
RPS is dynamic along two axes:
\begin{itemize}
  \item \textbf{Leaderboard refresh.} When new strong systems are added, $\mathrm{Best}_t$ is updated by rerunning the released evaluation scripts, which recalibrates all models' RPS accordingly.
  \item \textbf{Task expansion.} New tasks can be incorporated by contributing standardized evaluation scripts, allowing RPS to summarize a broader task set over time.
\end{itemize}
To support fair interpretation under heterogeneous task scopes, we report task-wise metrics alongside RPS summaries.

%% file: src/tracks.tex
\section{Track I: Scenario Stress Testing for Front-end Speech Tasks}
\label{sec:track1}

Track~I targets front-end speech perception, focusing on whether ASR systems remain reliable under realistic deployment conditions. We introduce this track because many evaluations are conducted on narrowly controlled test sets and therefore fail to characterize system behavior under the compound stressors that dominate practical deployments.

Our stress tests cover two complementary scenario families. First, we evaluate challenging acoustic and scenario conditions that frequently trigger recognition breakdowns, including background noise, far-field reverberation, and multi-speaker meeting speech. Concretely, we use VoxPopuli-en for naturally noisy English recordings \cite{wang2021voxpopuli}, AISHELL-5 for Mandarin in-car speech with noise and reverberation \cite{dai2025aishell5}, and meeting corpora AMI and AliMeeting to assess speaker-attributed transcription in English and Mandarin, respectively \cite{mccowan2005ami,yu2022m2met}. Second, we evaluate linguistic and context-dependent conditions that require explicit biasing or contextual grounding, including Mandarin--English code-switching (CS-Dialogue) \cite{zhou2025csdialogue}, dialectal variation (KeSpeech) \cite{tang2021kespeech}, and contextual/hotword-sensitive recognition (ContextASR) \cite{wang2025contextasr}.

We benchmark both modern speech foundation models and widely used conventional baselines, covering open-source systems and commercial APIs. For ASR, we report WER for English and CER for Mandarin. For meeting transcription, we additionally report permutation-invariant cpWER and DER.

\begin{table}[t]
  \centering
  \caption{Track I: Speaker-aware ASR performance (DER and cpWER/cpCER $\downarrow$ in \%). ``--'' denotes not available. Collar is 0.}
  \label{tab:meeting_wer_cpwer}
  \renewcommand{\arraystretch}{1.1}
  \small

  \resizebox{\columnwidth}{!}{%
  \begin{tabular}{l l c c}
    \toprule
    \textbf{Model} & \textbf{Type} & \textbf{AMI} & \textbf{AliMeeting} \\
    \cmidrule(lr){3-3} \cmidrule(lr){4-4}
     &  & DER / cpWER & DER / cpCER \\
    \midrule
    Diarizen+DiCoW        & Cascaded & 30.21 / 17.26  & -- \\
    Sortformer+FireRedASR & Cascaded & --             & 33.22 / 41.92 \\
    VibeVoice-ASR         & E2E SLM   & 41.26 / 36.80  & 47.33 / 43.66 \\
    \bottomrule
  \end{tabular}%
  }
\end{table}

The experiments reveal two complementary failure modes that motivate \textsc{SURE}'s scenario-driven design.
First, for speaker-aware meeting transcription (Table~\ref{tab:meeting_wer_cpwer}), cascaded pipelines remain highly competitive compared to end-to-end systems such as VibeVoice-ASR \cite{peng2026vibevoiceasr}, highlighting the difficulty of meeting-style conditions where far-field acoustics, interference, and speaker attribution interact. This gap underscores that meeting scenarios are not simply a harder ASR setting, but a compound of \emph{acoustic} stressors and \emph{structural} requirements (speaker permutation and attribution) that call for dedicated evaluation beyond single-speaker benchmarks.

\newcolumntype{Y}{>{\centering\arraybackslash}X}

\begin{table*}[t]
  \centering
  \caption{Track~I: Front-end perception evaluation under scenario stress tests.
  Error rates are reported in \% (lower is better). We additionally report RPS (\(\uparrow\)), where the task-specific SOTA is taken as the best score within the same table (thus RPS=1).
  ``--'' denotes not available. For ContextASR, we report \emph{with} hotword injection (left) and
  \emph{without} hotword injection (right); RPS uses the left value.}
  \label{tab:track1_frontend_rps}

  \setlength{\tabcolsep}{3.5pt}
  \renewcommand{\arraystretch}{1.15}
  \scriptsize

  \begin{tabularx}{\textwidth}{l *{12}{Y}}
    \toprule
    \multirow{3}{*}{\textbf{Model}} &
    \multicolumn{2}{c}{\textbf{Codeswitch}} &
    \multicolumn{2}{c}{\textbf{Dialect}} &
    \multicolumn{2}{c}{\textbf{Noise}} &
    \multicolumn{2}{c}{\textbf{Noise/Reverb}} &
    \multicolumn{2}{c}{\textbf{Context (En)}} &
    \multicolumn{2}{c}{\textbf{Context (Zh)}} \\
    \cmidrule(lr){2-3}\cmidrule(lr){4-5}\cmidrule(lr){6-7}\cmidrule(lr){8-9}\cmidrule(lr){10-11}\cmidrule(lr){12-13}
    & \multicolumn{2}{c}{\scriptsize CS-Dialogue \ \texttt{MER}\(\downarrow\)} &
      \multicolumn{2}{c}{\scriptsize KeSpeech \ \texttt{CER}\(\downarrow\)} &
      \multicolumn{2}{c}{\scriptsize VoxPopuli-en \ \texttt{WER}\(\downarrow\)} &
      \multicolumn{2}{c}{\scriptsize AISHELL-5 \ \texttt{CER}\(\downarrow\)} &
      \multicolumn{2}{c}{\scriptsize ContextASR-En \ \texttt{WER}\(\downarrow\)} &
      \multicolumn{2}{c}{\scriptsize ContextASR-Zh \ \texttt{CER}\(\downarrow\)} \\
    & \textbf{Raw} & \textbf{RPS} &
      \textbf{Raw} & \textbf{RPS} &
      \textbf{Raw} & \textbf{RPS} &
      \textbf{Raw} & \textbf{RPS} &
      \textbf{Raw} & \textbf{RPS} &
      \textbf{Raw} & \textbf{RPS} \\
    \midrule

    SenseVoice-Small   & 7.52 & 0.93 & 12.46 & 0.31 & 12.50 & 0.54 & 38.63 & 0.64 & 14.52 & 0.24 & 6.44 & 0.39 \\
    Whisper-large-v3   & 15.91 & 0.44 & 30.65 & 0.12 & 12.62 & 0.53 & 45.11 & 0.55 & 8.37  & 0.41 & 8.29 & 0.30 \\
    Parakeet-en        & -- & -- & -- & -- & \textbf{6.72} & \textbf{1.00} & -- & -- & 8.67 & 0.40 & -- & -- \\
    Gemini-2.5pro      & 17.96 & 0.39 & 31.82 & 0.12 & 9.03 & 0.74 & 64.49 & 0.38 & \textbf{3.47}/7.38 & \textbf{1.00} & 2.78 & 0.90 \\
    Qwen3-ASR-1.7B     & \textbf{7.00} & \textbf{1.00} & 5.12  & 0.74 & 7.41 & 0.91 & 25.46 & 0.97 & 5.58 & 0.62 & \textbf{2.50} & \textbf{1.00} \\
    FireRedLLM-L-7B    & 7.44 & 0.94 & \textbf{3.81} & \textbf{1.00} & 11.87 & 0.57 & \textbf{24.74} & \textbf{1.00} & 8.01 & 0.43 & 2.78/6.33 & 0.73 \\
    Kimi-Audio         & 11.94 & 0.59 & 7.80  & 0.49 & 10.63 & 0.63 & 45.72 & 0.54 & 6.66/7.56 & 0.52 & 2.96/3.82 & 0.84 \\
    \bottomrule
  \end{tabularx}
\end{table*}

Second, the ASR stress-test suite (Table~\ref{tab:track1_frontend_rps}) shows clear trade-offs across stressor families: systems with stronger context/biasing capabilities tend to perform better on code-switching and contextual recognition, but are not universally dominant under severe acoustic degradation or dialectal variation. We also observe that applying SURE's unified normalization and scoring can materially change reported results. For example, on LibriSpeech, rerunning a representative system under our evaluation pipeline \textit{yields an RPS shift of about 0.3} compared to the number quoted in its report, highlighting the necessity of a unified script for fair comparison.

Together, these results highlight the value of \textsc{SURE} for model selection by providing scenario-separated diagnostics under a unified protocol.

\section{Track~II: Full-stack Speech Understanding Evaluation}
\label{sec:track2}
Following the scenario stress tests in Track~I, Track~II performs a \emph{horizontal} comparison across representative speech understanding tasks under a unified protocol. We benchmark strong systems across paradigms, including end-to-end Speech LLMs and a cascaded pipeline as a complementary reference, and evaluate them using the same prediction format, normalization, and scoring scripts. As summarized in Table~\ref{tab:track2_understanding_clean}, Track~II covers a broad task spectrum ranging from basic recognition and translation to paralinguistic and semantic understanding.

\renewcommand\theadfont{\bfseries}

\begin{table}[t]
  \centering
  \caption{\textit{Track~II: Horizontal comparison on speech understanding tasks.}
  All scores are in \%. For ASR, we report LibriSpeech \texttt{WER} (clean/other) and AISHELL-1 \texttt{CER} ($\downarrow$).
  GR, SER, and SLU are accuracy ($\uparrow$). S2TT reports character-level BLEU on CoVoST2~\cite{wang2021covost} En$\leftrightarrow$Zh ($\uparrow$).
  ``--'' denotes not available.}
  \label{tab:track2_understanding_clean}

  \setlength{\tabcolsep}{3.2pt}
  \renewcommand{\arraystretch}{1.08}
  \fontsize{7.2}{8.4}\selectfont

  \setcellgapes{1.2pt}\makegapedcells

  \begin{tabularx}{\columnwidth}{c c *{5}{>{\centering\arraybackslash}X}}
\toprule
\textbf{Task} & \textbf{Dataset} & \textbf{Pipeline} &
\textbf{Gemini} & \textbf{Qwen3} & \textbf{Kimi} & \textbf{Gemini} \\
& & &
\textbf{3.0pro} & \textbf{Omni} & \textbf{Audio} & \textbf{2.5pro} \\
\midrule

    \multirow{2}{*}{\textbf{ASR}($\downarrow$)} &
    LibriSpeech &
    \makecell{2.90\\5.10} &
    \makecell{2.78\\4.40} &
    \makecell{\textbf{1.70}\\\textbf{3.05}} &
    \makecell{2.30\\3.83} &
    \makecell{3.07\\4.93} \\
    & AISHELL-1 &
    5.93 & 3.60 & 1.02 & \textbf{0.80} & 4.49 \\
    \midrule

    \textbf{GR}($\uparrow$) &
    LibriSpeech &
    53.69 & 78.50 & 82.74 & \textbf{92.02} & 59.64 \\
    \midrule

    \multirow{2}{*}{\textbf{S2TT}($\uparrow$)} &
    CoVoST2 En2Zh &
    18.12 & 15.92 & \textbf{46.25} & -- & 41.44 \\
    & CoVoST2 Zh2En &
    53.37 & 15.50 & 50.61 & -- & \textbf{60.14} \\
    \midrule

    \textbf{SER}($\uparrow$) &
    IEMOCAP &
    52.62 & 66.56 & 66.16 & \textbf{69.38} & 63.01 \\
    \midrule

    \textbf{SLU}($\uparrow$) &
    MMSU-Reason &
    76.45 & \textbf{89.07} & 83.61 & 75.33 & 84.64 \\
    \bottomrule
  \end{tabularx}
\end{table}

Three observations stand out. First, with fixed post-processing and scoring, cascaded pipelines can remain competitive on core perception tasks, indicating that a strong front-end coupled with a robust language back-end is still a viable design point under clean conditions. Second, emotion recognition remains challenging across all systems, suggesting that current models still under-exploit affective and prosodic cues. Third, we observe evaluation-critical \emph{format adherence} issues in some instruction-following Speech LLMs on relatively simple tasks such as ASR and S2TT: deviations from the required output schema can substantially degrade automatic metrics even when the generated content appears plausible.

\section{Track III: Initial Exploration of Controlled Training}
\label{sec:track3}

After Tracks~I--II, Track~III provides an initial exploration of \emph{controlled training} as a step toward more reproducible training-based studies. Rather than claiming broad architecture-level conclusions, our goal is to make ``paper + code'' \emph{executable and comparable} under a unified protocol by releasing an agent-assisted conversion workflow that ports training pipelines into the open-source framework \texttt{swift}~\cite{zhao2025swift}. Under a constrained open-data budget, we train models from scratch with a matched protocol and evaluate them at their best checkpoint using the same scoring scripts, reducing variance introduced by heterogeneous training pipelines and reporting.

\noindent\textbf{Tasks and data splits.}
We reuse the task spectrum of Track~II while constructing training splits that are source-related to the evaluation benchmarks and include explicit generalization checks. For example, we train SER on IEMOCAP~\cite{busso2008iemocap} and evaluate on MELD~\cite{poria2019meld}, and train SLU on SLURP~\cite{bastianelli2020slurp} while evaluating on MMSU-Reason~\cite{wang2025mmsu}. All metrics follow Track~II.

\begin{table}[t]
  \centering
  \caption{Track~III task coverage for controlled training. ASR is evaluated on Aishell1(Zh) and LibriSpeech test-clean(En); GR on LibriSpeech; SER on MELD; SLU on MMSU-Reason; S2TT on CoVoST2. All metrics follow Table~\ref{tab:track2_understanding_clean} and are reported in \%.}
  \label{tab:track3_task_coverage}
  \setlength{\tabcolsep}{4pt}
  \renewcommand{\arraystretch}{1.2}
  \small
  \resizebox{\columnwidth}{!}{%
    \begin{tabular}{l c c c c c}
      \toprule
      \multirow{2}{*}{\textbf{Model}} &
      \multirow{2}{*}{\textbf{ASR}$\downarrow$} &
      \multirow{2}{*}{\textbf{GR}$\uparrow$} &
      \multirow{2}{*}{\textbf{S2TT}$\uparrow$} &
      \multirow{2}{*}{\textbf{SER}$\uparrow$} &
      \multirow{2}{*}{\textbf{SLU}$\uparrow$} \\
      & \scriptsize{Zh/En} & & \scriptsize{En$\rightarrow$Zh / Zh$\rightarrow$En} & & \\
      \midrule
      Qwen2-audio~\cite{chu2024qwen2} & 1.58/2.57 & 98.93 & 33.00/43.36 & 40.38 & 47.81 \\
      TASU~\cite{peng2026tasutextonlyalignmentspeech}     & 4.36/3.30 & 46.78 & 32.41/34.54 & 31.49 & 45.13 \\
      \bottomrule
    \end{tabular}%
  }
\end{table}

\vspace{0.25em}
\noindent\textbf{Agent-assisted conversion workflow.}
As shown on the right of Fig.~\ref{fig:sure_structure}, our agent pipeline produces a \texttt{swift} training recipe together with a versioned conversion plan and validation reports. It analyzes model specifications from papers and repositories, generates an executable configuration, verifies data and loss/metric wiring, and runs integration checks before launching training. Concretely, the agent materializes (i) a versioned \texttt{swift} recipe (model, data, optimizer, and schedules), (ii) an executable conversion plan, and (iii) validator reports. The validator performs static checks (dependency resolution, config sanity, loss/metric signatures) and integration checks (a short smoke-run on a small batch) to ensure the converted pipeline is runnable before full training.

\vspace{0.25em}
\noindent\textbf{Initial model coverage and results.}
As a proof of concept, we port a small set of representative models into \texttt{swift}. Notably, \emph{Qwen2-Audio} can be converted end-to-end without manual patches, while other models may require lightweight human edits due to incomplete or non-standard releases. Table~\ref{tab:track3_task_coverage} reports results for \emph{Qwen2-Audio-7B} and \emph{TASU(SFT)-2B}, both trained from scratch under the same protocol. Overall, \emph{TASU} lags behind \emph{Qwen2-Audio} on paralinguistic tasks (e.g., GR and SER), while remaining competitive on semantic tasks (e.g., SLU and S2TT), which is consistent with its design emphasis on language-oriented supervision.

%% file: src/conclusion.tex
\section{Conclusions}
We presented \textsc{SURE}, a unified and reproducible experimentation framework for speech understanding. \textsc{SURE} standardizes prediction formats, normalization, and scoring for consistent comparison across model types, and provides scenario-driven suites under realistic acoustic and linguistic stressors. It also releases an agent-assisted conversion workflow that turns ``paper + code'' into versioned, runnable \texttt{swift} pipelines for controlled training studies. \textsc{SURE} is open-sourced and extensible for deployment-oriented model selection.